\documentclass[runningheads]{llncs}
\usepackage[T1]{fontenc}
\usepackage{graphicx}
\usepackage{booktabs}
\usepackage{amsmath}
\begin{document}
\title{Sales Skills Training in Virtual Reality: An evaluation utilizing CAVE and Virtual Avatars}
\titlerunning{Sales Skills Training in Virtual Reality}
%
\author{Francesco Vona\inst{1}\orcidID{0000-0003-4558-4989} \and
Michael Stern\inst{1}\orcidID{0009-0002-7558-5384} \and
Navid Ashrafi\inst{1}\orcidID{0009-0005-8398-415X} \and
Julia Schorlemmer\inst{1}\orcidID{0009-0004-7388-9389} \and
Jessica Stemann\inst{1}\orcidID{0000-0002-0361-6754} \and
Jan-Niklas Voigt-Antons\inst{1}\orcidID{0000-0002-2786-9262} \
}
\authorrunning{F. Vona et al.}
%
\institute{University of Applied Sciences Hamm-Lippstadt \\
\email{name.lastname@hshl.de}}
\maketitle              
\begin{abstract}
This study investigates the potential of virtual reality (VR) for enhancing sales skills training using a Cave Automatic Virtual Environment (CAVE). VR technology enables users to practice interpersonal and negotiation skills in controlled, immersive environments that mimic real-world scenarios. In this study, participants engaged in sales simulations set in a virtual dealership, interacting with avatars in different work settings and with various communication styles.
The research employed a within-subjects experimental design involving 20 university students. Each participant experienced four distinct sales scenarios randomized for environmental and customer conditions. Training effectiveness was assessed using validated metrics alongside custom experience questions.
Findings revealed consistent user experience and presence across all scenarios, with no significant differences detected based on communication styles or environmental conditions. The study highlights the advantages of semi-immersive VR systems for collaborative learning, peer feedback, and realistic training environments. However, further research is recommended to refine VR designs, improve engagement, and maximize skills transfer to real-world applications.

\end{abstract}
\section{Introduction}
Sales and negotiation skills are fundamental to business success \cite{1}, and their importance is particularly important in the context of a highly competitive global market and always rising customer expectations \cite{2}. Traditionally, sales training relies heavily on the development of interpersonal skills through role-playing exercises. In these exercises, the trainers act as customers, allowing the trainees to practice in a low-stakes environment. This approach, known as face-to-face training, requires significant resources, as it depends on the expertise of training specialists. Consequently, face-to-face training is often the most expensive form of traditional training \cite{2}.
Despite its high cost, face-to-face training remains widely used, often supplemented with paper-based or video-based materials \cite{5}. However, traditional methods like role-plays and business simulations depend heavily on participant performance and adaptability, which can limit their overall effectiveness. These limitations highlight the need for innovative training approaches to provide experiential, interactive, and scalable learning opportunities in sales and negotiation.
Immersive media, including virtual reality (VR) and, more generally, extended reality (XR), has emerged as a promising alternative to traditional methods. Technologies like head-mounted displays and cave automatic virtual environments (CAVE) create realistic, engaging, and interactive training scenarios, addressing many limitations of conventional techniques \cite{2}. Unlike static training materials, VR can replace traditional paper-based questionnaires with more immersive VR assessments, offering a more engaging way to measure trainee performance and learning outcomes \cite{9a}. Additionally, VR and virtual environments can induce emotional states more effectively than other media types, which is particularly beneficial for realistic and impactful training scenarios \cite{12voi}. 

In particular, VR offers unique advantages for sales training. It allows trainees to practice in lifelike simulations that closely mimic real-world scenarios, leading to improved skill retention and application in actual job settings \cite{2}. Unlike traditional role-play exercises, VR enables risk-free learning environments where trainees can make mistakes and refine their skills without real-world consequences \cite{9}. Moreover, VR-based training can standardize learning experiences, ensuring consistent quality across participants, while providing immediate feedback.

CAVEs, as an advanced XR tool, provide additional benefits for immersive training \cite{17}. They create collaborative, large-scale virtual environments where trainees can practice negotiation and interpersonal skills in dynamic, team-based settings. These technologies expand the scope of experiential learning, offering structured, interactive simulations that facilitate the development of critical sales and negotiation competencies \cite{7}.
Despite the growing interest in immersive media for education, there is a lack of empirical studies focusing on the development, testing, and evaluation of immersive training tools tailored specifically for sales and negotiation contexts. Current research has largely overlooked the potential of technologies like CAVEs in addressing the shortcomings of traditional training approaches.

This study seeks to fill this gap by exploring the application of CAVEs in simulations for sales and negotiation training. By focusing on the unique capabilities of CAVE environments, this research aims to create a scalable, engaging, and effective platform for immersive learning.
The primary goal of this study is to harness the potential of immersive media, particularly CAVEs, to enhance sales training. Specifically, the objectives include: i) Designing a user-centered sales and negotiation simulation tailored for CAVE environments, ii) implementing this simulation to create an interactive and engaging learning experience, and iii) conducting an initial usability study to evaluate the effectiveness of the CAVE-based training approach.

\section{Related Work}
Virtual reality training programs have demonstrated success across diverse domains, including gamified shooting simulations \cite{8}, sommelier training \cite{9}, and technical skill development \cite{10}. Increasingly, VR training is being adopted in educational contexts, offering immersive and engaging experiences that improve outcomes compared to traditional methods \cite{11,14,13}. The immersive nature of VR enhances realism and engagement, providing trainees with dynamic environments to practice and refine skills. These features are particularly valuable in sales training, where interpersonal skills are critical \cite{2,13}.
VR enables realistic simulations of job scenarios, allowing trainees to practice in a risk-free environment. This immersive experience has been shown to improve performance in actual workplace settings, producing outcomes superior to those achieved through traditional alternatives \cite{14}. For instance, research  \cite{13} highlights the application of VR in sales training, using case studies and cutting-edge research to explore its implications for practice. However, despite these advancements, there remains a significant gap in the development of VR systems dedicated specifically to sales training, emphasizing the need for further research in this area \cite{2}.

CAVE systems have also been successfully applied in various fields, including education, safety training, and engineering \cite{20,19,18,21}. Their interactive and immersive nature has been shown to enhance learning experiences, making them versatile tools in modern training environments \cite{17,16,15}. However, the effectiveness of CAVE varies depending on the application and individual learner abilities, underscoring the importance of ongoing research and development in this domain.
In education, CAVE systems have been used to teach fire safety skills to children through game-like interactions, increasing engagement and motivation. These systems make standard safety information more engaging and enjoyable, leading to improved learning outcomes \cite{18}. Similarly, serious games for school fire prevention have leveraged CAVE to provide realistic, interactive simulations, enhancing learning through hands-on discovery \cite{19}.
CAVE has also been explored in the context of emotional intelligence training. While it can simulate emotional scenarios effectively, studies indicate that training success often depends on trainees’ spatial abilities rather than emotional intelligence itself. This suggests that, while VR provides valuable simulations, it may not fully replicate complex interpersonal communication \cite{20}. In sports, CAVE systems have been used to train athletes for high-pressure scenarios by inducing controlled anxiety. While promising, further research is needed to evaluate the long-term benefits of these applications \cite{21}.
Last, CAVE was also explored in engineering education, where it was compared with other VR setups, demonstrating superior outcomes in student achievement. The immersive and interactive nature of CAVE provides a more engaging learning experience, resulting in better educational outcomes than traditional methods \cite{22}.

This work builds upon the authors’ earlier research on user-centered simulations for leadership development in CAVE environments \cite{12}. In their previous study, the authors designed and evaluated leadership training scenarios tailored to simulate realistic workplace situations, such as providing critical feedback or addressing health concerns. These scenarios enabled participants to interact with virtual characters in dynamic, context-rich environments, delivering high levels of user presence and interactivity. The findings underscored the potential of CAVE as an effective tool for experiential learning, particularly in domains requiring interpersonal skill development, such as leadership and sales training.

\section{Methods}
\subsection{Study Design}
The study employed a within-subjects experimental design to ensure that all participants experienced each of the testing conditions. The goal was to observe and assess how participants adapted their communication strategies and perceived their experiences across different contexts. The order of the conditions was randomized, minimizing potential biases related to individual differences. The experimental conditions were designed to simulate real-world sales scenarios in a controlled virtual environment. 
Participants engaged in role-playing exercises, interacting with virtual customers who displayed varying personalities and communication styles in different working environments. The four distinct scenarios combined two factors: Customer (Avatar) Personality (Friendly vs. Unfriendly) and Environmental Atmosphere (Friendly vs. Unfriendly). Conditions were defined as follows: Condition 1 = Friendly User x Friendly Environment (FUxFE), Condition 2 = Friendly User x Unfriendly Environment (FUxUE), Condition 3 = Unfriendly User x Friendly Environment (UUxFE), and Condition 4 = Unfriendly User x Unfriendly Environment (UUxUE). This factorial design allowed researchers to systematically explore how these variables influenced participants’ user experiences and interactions. 
\begin{figure}[ht!]
    \centering    \includegraphics[width=\textwidth]{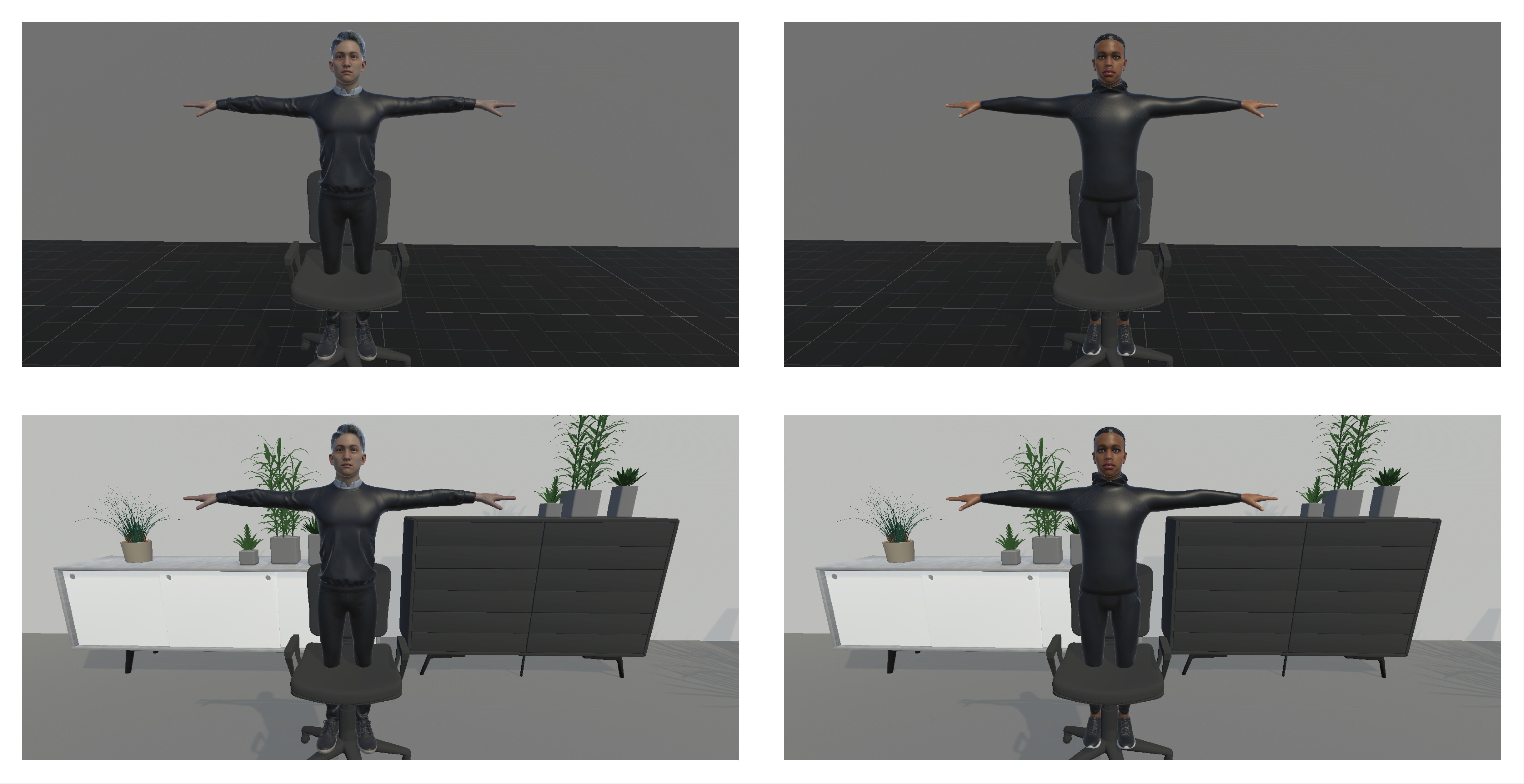}
    \caption{The four conditions that were tested during the experiment: Condition 1 = Friendly User x Friendly Environment (bottom left), Condition 2 = Friendly User x Unfriendly Environment (top left), Condition 3 = Unfriendly User x Friendly Environment (bottom right), and Condition 4 = Unfriendly User x Unfriendly Environment (top right).}
    \label{fig:conditions}
\end{figure}
In each scenario, the participants had to follow a pre-configured script. Figure \ref{fig:conditions} provides a visual representation of the conditions tested during the study.

\subsection{Participants}
Participants were recruited from the student population of our university. The final sample comprised 20 individuals, with an average age of 24.65 years (SD = 4.20). The sample included 75\% male participants (n = 15) and 25\% female participants (n = 5). Participants' affinity for technology interaction (ATI) was assessed using the ATI scale \cite{ati}, with an overall mean score of 3.65 (\textit{SD} = 1.09), indicating moderate comfort and familiarity with technology within the sample. Recruitment efforts ensured diversity in educational backgrounds, though all participants had some level of familiarity with sales concepts, either through coursework or extracurricular activities. Inclusion criteria required participants to have no prior experience with the specific VR system used in the study. To encourage participation and ensure adequate representation, participants who were not employed by the host institution received a monetary compensation of 15 Euros per hour for their time and effort. The study was conducted in compliance with ethical guidelines and received approval from the university’s local ethics commission.

\subsection{Measures}
A combination of quantitative and qualitative measures was used to evaluate participants’ experiences:
\begin{itemize}
    \item Presence: The Igroup Presence Questionnaire (IPQ) was administered to measure participants’ sense of being “present” in the virtual environment. This scale assessed factors such as spatial presence, involvement, and realism \cite{ipq}. 
    \item Affinity for Technology Interaction: The ATI (Affinity for Technology Interaction) scale was used to profile participants’ general attitude and comfort with technology. This measure provided insights into individual differences that could influence user experience in a VR setting \cite{ati}. 
    \item User Experience: The User Experience Questionnaire - Short Version (UEQ-S) was employed to evaluate participants’ overall satisfaction with the VR environment. The scale captured dimensions such as pragmatic and hedonic quality, and usability \cite{ueqs}.
    \item Social Presence: The Social Presence Questionnaire (SPQ) measured participants' perceived social presence, focusing on the sense of being with and interacting with others in virtual environments. The SPQ evaluates dimensions such as mutual awareness, co-presence, and interaction quality \cite{spq}.
    \item Custom Questions: A set of custom questions was included after each scenario to gather additional feedback on the specific interaction and perceived challenges. These questions were designed to identify contextual nuances not captured by standardized measures. The custom questions were assessed on a 7-point Likert scale. The items are as follows: CUSQ1: "How realistic did the interview situation feel?" (1 = Not realistic at all, 7 = Very realistic); CUSQ2: "How well did you feel during the simulated interview?" (1 = Bad, 7 = Very good); CUSQ3: "Did you experience any discomfort or pain during the interview?" (1 = No discomfort at all, 7 = A lot of discomfort); CUSQ4: "Did you achieve your goal for this interview?" (1 = Not achieved at all, 7 = Fully achieved); CUSQ5: "Did you experience any challenges during the interview?" (1 = No challenges at all, 7 = A lot of challenges).
\end{itemize}
At the end of the session, participants were allowed to leave further feedback about their overall experience.
\subsection{Procedure}
\begin{figure}[ht!]
    \centering\includegraphics[width=\textwidth]{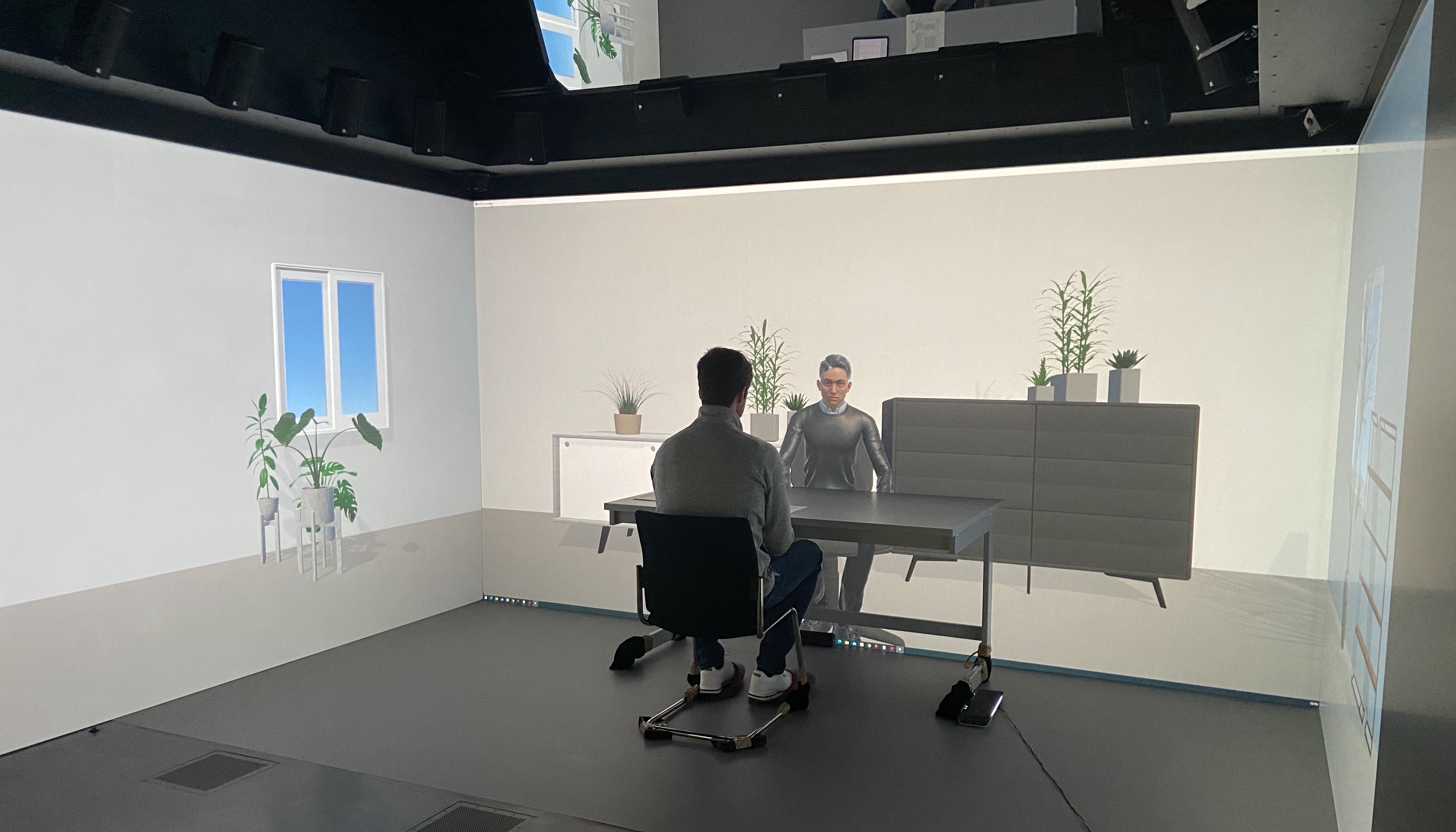}
    \caption{A participant was photographed during the study, sitting on a chair in the CAVE system.  Between him and the avatar, a table was placed to enhance the realism of the setting.}
    \label{fig:experiment}
\end{figure}

Upon arrival, participants were welcomed and briefed on the study’s purpose and procedure. After signing an informed consent form, they were asked to complete a demographics questionnaire to collect information about their age, gender, educational background, and prior experience with VR technology or sales scenarios.
The CAVE system was initialized, and the physical room was arranged to resemble a dealership office. This setup included realistic props like a desk and chairs to create a contextually relevant and engaging environment. Participants were then introduced to the sales task and given a general briefing about the sales scenarios.
Each participant completed four scenarios, with each scenario lasting approximately five minutes. In these scenarios, participants assumed the role of a salesperson tasked with selling either a used car or motorcycle to a virtual customer. The customer’s personality (friendly vs. unfriendly) and environmental atmosphere (friendly vs. unfriendly) varied across scenarios, ensuring exposure to all conditions. To maintain consistency, participants were provided with conversation guidelines to structure their interactions (Figure \ref{fig:experiment}).
After completing each scenario, participants filled out a post-run questionnaire to evaluate their experience and assess their interaction with the virtual customer.
Upon completing all scenarios, participants were asked to fill out a final questionnaire summarizing their overall experience. This questionnaire included both quantitative measures (e.g., standardized scales) and open-ended questions to capture qualitative insights. The entire study session, including briefing, scenarios, and debriefing, lasted approximately 60 minutes per participant.
\section{Results}
A series of analyses were conducted to evaluate the participants’ experiences and interactions within the simulated sales scenarios. Initially, descriptive statistics were computed to summarize the data and provide an overview of the participants’ responses across the different conditions. Subsequently, repeated-measures analyses of variance (RM-ANOVAs) were performed to examine potential differences between the experimental conditions concerning the user experience, the sense of presence, and social presence. However, these analyses did not reveal any statistically significant effects. Following this, additional qualitative analyses were carried out to explore the quantitative data in greater depth, aiming to identify trends and insights that could further inform the study's findings.

\subsection{Descriptive statistics}

Descriptive statistics were calculated for all measured variables to provide an overview of participants' experiences across different conditions. These include the User Experience Questionnaire - Short Version (UEQ-S), the Igroup Presence Questionnaire (IPQ), the Social Presence Questionnaire (SPQ), and a set of custom questions tailored to assess specific aspects of the simulated interviews. The descriptive analysis highlights the average (\textit{M}) and variability (\textit{SD}) within each condition, offering insights into participants' perceptions of usability, presence, and various qualitative aspects of the interview process.

\subsubsection{User Experience Questionnaire (UEQ-S).}
The descriptive analysis of the UEQ-S revealed variations in both pragmatic and hedonic quality across the four conditions. For Pragmatic Quality, mean scores were highest in Condition 2 (FUxUE) (\textit{M} = 4.65, \textit{SD} = 1.22), followed closely by Condition 1 (FUxFE) (\textit{M} = 4.60, \textit{SD} = 1.42) and Condition 3 (UUxFE) (\textit{M} = 4.60, \textit{SD} = 1.35). Condition 4 (UUxUE) showed the lowest mean score (\textit{M} = 4.43, \textit{SD} = 1.52), indicating reduced pragmatic usability in an unfriendly user and environment context.
For Hedonic Quality, the highest mean score was observed in Condition 2 (FUxUE) (\textit{M} = 4.43, \textit{SD} = 1.13), followed by Condition 1 (FUxFE) (\textit{M} = 4.39, \textit{SD} = 1.38). Scores decreased in Condition 3 (UUxFE) (\textit{M} = 4.26, \textit{SD} = 1.46) and Condition 4 (UUxUE) (\textit{M} = 4.14, \textit{SD} = 1.52), reflecting a diminished sense of enjoyment and engagement in less favorable scenarios.

\subsubsection{Igroup Presence Questionnaire (IPQ).}
The total scores of the IPQ demonstrated consistent perceptions of presence across conditions, with mean values spanning from 3.17 to 3.33. The highest presence score was recorded in Condition 4 (UUxUE) (\textit{M} = 3.33, \textit{SD} = 1.15), while the lowest was observed in Condition 2 (FUxUE) (\textit{M} = 3.17, \textit{SD} = 1.10). These findings suggest that participants’ sense of presence remained relatively stable across scenarios, with minor variations. It is interesting to note that the feeling of presence does not decrease with "unfriendly" conditions.

\subsubsection{Social Presence Questionnaire (SPQ).}
Descriptive statistics for the SPQ showed minor differences in perceived social presence. The highest mean score was observed in Condition 2 (FUxUE) (\textit{M} = 3.29, \textit{SD} = 1.06), while the lowest was in Condition 1 (FUxFE) (\textit{M} = 3.18, \textit{SD} = 1.14). Condition 4 (UUxUE) (\textit{M} = 3.26, \textit{SD} = 1.21) and Condition 3 (UUxFE) (\textit{M} = 3.19, \textit{SD} = 1.12) demonstrated slightly higher perceived social presence scores compared to Condition 1.

\bigskip Table \ref{tab:descriptive_stats} presents the descriptive statistics, including the means and standard deviations, for the User Experience Questionnaire, the Igroup Presence Questionnaire, and the Social Presence Questionnaire across all four conditions.

\begin{table}[ht]
\centering
\caption{Descriptive Statistics for UEQ-S, IPQ, and SPQ across conditions. The total values represent aggregated mean scores across all items within each respective questionnaire.}
\label{tab:descriptive_stats}
\begin{tabular}{lcccc}
\toprule
\textbf{Variable} & \textbf{Condition 1} & \textbf{Condition 2} & \textbf{Condition 3} & \textbf{Condition 4} \\
& \textit{M (SD)} & \textit{M (SD)} & \textit{M (SD)} & \textit{M (SD)} \\
\midrule
\textbf{Pragmatic Quality (UEQ-S)} & 4.60 (1.42) & 4.65 (1.22) & 4.60 (1.35) & 4.43 (1.52) \\
\textbf{Hedonic Quality (UEQ-S)} & 4.39 (1.38) & 4.43 (1.13) & 4.26 (1.46) & 4.14 (1.52) \\
\textbf{UEQ-S total value} & 4.50 (1.30) & 4.54 (1.18) & 4.43 (1.40) & 4.28 (1.50) \\
\textbf{IPQ total value} & 3.25 (1.06) & 3.17 (1.10) & 3.31 (1.02) & 3.33 (1.15) \\
\textbf{SPQ total value} & 3.18 (1.14) & 3.29 (1.06) & 3.19 (1.12) & 3.26 (1.21) \\
\bottomrule
\end{tabular}
\end{table}

\subsubsection{Custom questions (CUSQ).}
The custom questions assessed participants’ perceptions of the interview situation across four conditions, focusing on realism, well-being, discomfort, goal achievement, and challenges experienced.

CUSQ1 (Realism): Participants rated how realistic the interview situation felt. Across the four conditions, ratings ranged from moderately to highly realistic. The highest realism score was observed in Condition 4 (UUxUE) (\textit{M} = 3.45, \textit{SD} = 1.36), while Condition 1 (FUxFE) scored the lowest (\textit{M} = 2.90, \textit{SD} = 1.33).

CUSQ2 (Well-being): Ratings for participants’ well-being during the simulated interview varied slightly across conditions. Condition 4 (UUxUE) yielded the highest average score (\textit{M} = 4.75, \textit{SD} = 1.62), suggesting participants felt best in this scenario. Conversely, Condition 1 (FUxFE) showed the lowest mean score (\textit{M} = 4.00, \textit{SD} = 1.62).

CUSQ3 (Discomfort): Participants reported their levels of discomfort or pain during the interview. Scores remained relatively low across all conditions, with Condition 3 (UUxFE) showing the lowest average discomfort (\textit{M} = 3.25, \textit{SD} = 1.48) and Condition 4 (UUxUE) the highest (\textit{M} = 3.45, \textit{SD} = 1.36).

CUSQ4 (Goal Achievement): Participants assessed whether they achieved their goals during the interview. Scores were fairly consistent, with Condition 2 (FUxUE) having the highest mean score (\textit{M} = 4.75, \textit{SD} = 1.62), indicating greater perceived goal achievement. The lowest score was observed in Condition 1 (FUxFE) (\textit{M} = 4.00, \textit{SD} = 1.62).

CUSQ5 (Challenges): Finally, participants rated the extent of challenges experienced during the interview. Scores were comparable across conditions, with Condition 3 (UUxFE) showing slightly higher levels of challenges (\textit{M} = 3.25, \textit{SD} = 1.48) compared to Condition 4 (UUxUE), which had the lowest (\textit{M} = 3.00, \textit{SD} = 1.62).

Table \ref{tab:cusq_stats} presents the descriptive statistics, including the means and standard deviations, for the custom questions.

\begin{table}[ht]
\centering
\caption{Descriptive Statistics for Custom Questions Across Conditions. The custom questions were assessed on a 7-point Likert scale.}
\label{tab:cusq_stats}
\begin{tabular}{lcccc}
\toprule
\textbf{Custom Question} & \textbf{Condition 1} & \textbf{Condition 2} & \textbf{Condition 3} & \textbf{Condition 4} \\
& \textit{M (SD)} & \textit{M (SD)} & \textit{M (SD)} & \textit{M (SD)} \\
\midrule
\textbf{CUSQ1 (Realism)} & 2.90 (1.33) & 3.20 (1.40) & 3.25 (1.48) & 3.45 (1.36) \\
\textbf{CUSQ2 (Well-being)} & 4.00 (1.62) & 4.50 (1.40) & 4.60 (1.48) & 4.75 (1.62) \\
\textbf{CUSQ3 (Discomfort)} & 3.00 (1.33) & 3.20 (1.40) & 3.25 (1.48) & 3.45 (1.36) \\
\textbf{CUSQ4 (Goal Achievement)} & 4.00 (1.62) & 4.75 (1.62) & 4.60 (1.48) & 4.50 (1.40) \\
\textbf{CUSQ5 (Challenges)} & 3.00 (1.33) & 3.20 (1.40) & 3.25 (1.48) & 3.00 (1.62) \\
\bottomrule
\end{tabular}
\end{table}

\subsection{Repeated-measures ANOVA}
To examine differences across the experimental conditions, repeated-measures ANOVAs were conducted for each dependent variable: Pragmatic Quality (UEQ-S), Hedonic Quality (UEQ-S), UEQ-S Total Value, IPQ Total Value, and SPQ Total Value. While none of the analyses revealed statistically significant differences between conditions, these results are presented to ensure transparency and methodological rigor.

Including these findings allows a comprehensive understanding of the data and ensures that even non-significant outcomes are documented. Furthermore, the effect size measures ($\eta^2_{\text{G}}$) provide insights into the magnitude of the observed effects, which may inform future research or guide experimental design adjustments.

\subsubsection{Repeated-Measures ANOVA: Pragmatic Quality (UEQ-S).}

A repeated-measures ANOVA was conducted to examine the effect of condition on pragmatic quality, as assessed by the UEQ-S. The analysis revealed no significant main effect of condition, \textit{F}(3, 57) = 0.24, \textit{p} = 0.865, $\eta^2_{\text{G}} = 0.004$. Mauchly’s test indicated that the assumption of sphericity was met, W=0.81, \textit{p} = 0.588. Consequently, no sphericity corrections were applied.

Post-hoc pairwise comparisons using Bonferroni adjustments revealed no significant differences between any pair of conditions (\textit{p} > 0.05).

\subsubsection{Repeated-Measures ANOVA: Hedonic Quality (UEQ-S).}

The analysis for hedonic quality, measured by the UEQ-S, indicated no significant differences between conditions, \textit{F}(3, 57) = 0.81, \textit{p} = 0.495, $\eta^2_{\text{G}} = 0.007$. Mauchly’s test confirmed that the sphericity assumption was not violated (W=0.74, \textit{p} = 0.375), and sphericity corrections were therefore unnecessary.

Pairwise comparisons with Bonferroni adjustments did not reveal significant differences across conditions (\textit{p} > 0.05).

\subsubsection{Repeated-Measures ANOVA: Total Value of UEQ-S.}

A repeated-measures ANOVA was conducted to evaluate the effect of the condition on the total value of the UEQ-S. The analysis revealed no significant main effect of condition, \textit{F}(3, 57) = 0.64, \textit{p} = 0.594, $\eta^2_{\text{G}} = 0.007$. Mauchly’s test indicated that the assumption of sphericity was met (W=0.72, \textit{p} = 0.319), and sphericity corrections were therefore unnecessary.

Post-hoc pairwise comparisons using Bonferroni adjustments showed no significant differences between any pair of conditions (\textit{p} > 0.05).

\subsubsection{Repeated-Measures ANOVA: Total Value of IPQ.}

The repeated-measures ANOVA assessing the impact of condition on the total value of the IPQ revealed no significant main effect of condition, \textit{F}(3, 57) = 0.22, \textit{p} = 0.879, $\eta^2_{\text{G}} = 0.004$. Mauchly’s test indicated a violation of the sphericity assumption (W=0.39W=0.39, \textit{p} = 0.005). However, sphericity corrections using Greenhouse-Geisser (\textit{p} = 0.815) and Huynh-Feldt (\textit{p} = 0.840) estimates did not alter the non-significant result.

Post-hoc pairwise comparisons with Bonferroni adjustments revealed no significant differences between any pair of conditions (\textit{p} > 0.05).

\subsubsection{Repeated-Measures ANOVA: Total Value of SPQ.}

The analysis for the total value of the SPQ showed no significant main effect of condition, \textit{F}(3, 57) = 0.21, \textit{p} = 0.887, $\eta^2_{\text{G}} = 0.002$. Mauchly’s test confirmed that the assumption of sphericity was not violated (W=0.69W=0.69, \textit{p} = 0.246), so no corrections were applied.

Post-hoc pairwise comparisons using Bonferroni adjustments revealed no significant differences between any pair of conditions (\textit{p} > 0.05).

\subsection{Qualitative Results}
Qualitative feedback was collected to provide deeper insights into participants' experiences during the simulations. Participants were asked open-ended questions about the virtual environment, interactions with avatars, challenges faced, and areas for improvement. The responses were analyzed thematically, highlighting the critical aspects of their experiences.
\subsubsection{Environment.} 
Participants’ perceptions of the simulated environment varied significantly. Many appreciated the immersive quality of certain conditions, with one participant noting, \emph{"The friendly environment felt welcoming and made the tasks more manageable."} However, unfriendly conditions were described as "cold" and "distracting," with participants reporting that the visual and auditory elements were sometimes exaggerated, reducing realism. Some mentioned inconsistencies, such as static objects or limited interactivity, which detracted from the overall experience.
\subsubsection{Interaction with Avatars.}
Feedback on avatar interactions was similarly mixed. Participants appreciated that avatars introduced a human element into the simulation, with one stating, \emph{"The avatars helped simulate real conversations, which was engaging."} However, others noted that the avatars' behavior sometimes felt "robotic" or "repetitive," with limited adaptability to user input. Participants suggested enhancing the avatars' responsiveness and increasing the variety in their communication styles to make interactions more dynamic and realistic.
\subsubsection{Challenges encountered.}
Participants reported a range of challenges, primarily related to technical issues and task complexity. For example, lag and delayed responses occasionally disrupted the flow of tasks, with one participant mentioning, \emph{"The system froze briefly, which broke my concentration."} Additionally, some found the cognitive load overwhelming, particularly in conditions with both unfriendly users and environments, describing it as "stressful to the point of distraction."
\subsubsection{Suggestions for Improvement.} 
Participants proposed several improvements to enhance the simulation experience. Many emphasized the need for greater realism in environmental and avatar interactions. For example, one participant recommended, \emph{"Making objects in the environment respond to user actions would make the scenarios more realistic."} Others suggested optimizing system performance to minimize technical issues and introducing more varied scenarios to increase engagement and reflect real-world complexity.


\section{Discussion}
The present study aimed to evaluate participants’ experiences and interactions within simulated sales scenarios by examining user experience, presence, and social presence across different experimental conditions. Although the results did not yield statistically significant differences between conditions, the findings provide valuable insights into the nuances of participants’ experiences and suggest potential areas for improvement in future simulations.

\subsubsection{User Experience and Perceived Quality.}
The descriptive statistics for the User Experience Questionnaire indicated that pragmatic and hedonic quality ratings varied slightly across conditions. Conditions with friendly users (FUxUE and FUxFE) tended to yield higher scores for both pragmatic and hedonic quality, suggesting that a more welcoming and supportive context enhances usability and enjoyment. Interestingly, the lowest scores for both dimensions were observed in the condition with unfriendly users and an unfriendly environment (UUxUE), reinforcing the importance of fostering a positive interactional and environmental context to optimize user experience.

Despite these trends, the repeated-measures ANOVAs revealed no statistically significant effects of condition on pragmatic or hedonic quality. This finding may suggest that while users are sensitive to contextual variations, the overall impact of these variations on their experience may not be strong enough to produce measurable differences within the scope of this study. Alternatively, the measures employed or the sample size might not have been sufficient to detect small but meaningful effects.

\subsubsection{Presence and Social Presence.}

The Igroup Presence Questionnaire (IPQ) and Social Presence Questionnaire (SPQ) results showed stable perceptions of presence across conditions. Notably, the condition with unfriendly users and an unfriendly environment (UUxUE) did not result in diminished presence, contrary to expectations. This finding suggests that participants’ sense of “being there” and their perception of social presence may be more resilient to adverse contextual factors than previously assumed. However, qualitative feedback revealed that unfriendly conditions were often perceived as less immersive or realistic, highlighting a potential disconnect between quantitative measures of presence and participants’ subjective experiences.

\subsubsection{Custom Questions and Qualitative Insights.}

Custom questions provided additional insights into participants’ perceptions of realism, well-being, discomfort, goal achievement, and challenges across conditions. Realism and well-being scores were highest in the UUxUE condition, an unexpected finding given the unfriendly context. This result may reflect participants’ adaptation to challenging scenarios or a heightened sense of accomplishment in overcoming adversity. However, qualitative feedback highlighted that unfriendly conditions were sometimes described as “cold” or “distracting,” suggesting that while participants adapted, their experiences were not uniformly positive. Participants also identified challenges such as technical issues, high cognitive load, and limited interactivity in the simulations. These challenges were particularly pronounced in conditions involving both unfriendly users and environments, where participants reported feeling overwhelmed or distracted.

\subsubsection{Lesson learned}
Finally, the set of descriptive results and qualitative feedback was distilled into a set of lessons learned that can be useful for the development of similar scenarios:
\begin{itemize}
    \item \textbf{Context Matters:} Friendly environments and interactions appear to enhance user experience, suggesting that incorporating supportive and engaging elements can improve usability and enjoyment. However, designers should also consider how to make unfriendly scenarios realistic yet manageable, as they are often necessary for training purposes.
    \item \textbf{Enhancing Realism:} Participants emphasized the need for greater realism in both environmental and avatar interactions. Features such as dynamic object behavior, more responsive avatars, and varied communication styles could make simulations more engaging and reflective of real-world scenarios.
    \item \textbf{Minimizing Technical Issues:} Technical disruptions, such as lag or system freezes, were reported to disrupt participants’ focus and immersion. Optimizing system performance should be a priority to ensure a seamless experience.
    \item \textbf{Balancing Cognitive Load:} High task complexity and simultaneous challenges were reported as overwhelming in some conditions. Future simulations should aim to balance cognitive demand to maintain user engagement without inducing excessive stress.
\end{itemize}


\section{Conclusion}
To conclude, immersive media holds immense potential for experiential learning and training in sales and negotiation contexts. By offering realistic, and immersive experiences, platforms such as the VR CAVE provide learners with state-of-the-art training opportunities, overcoming barriers related to geographical, and time constraints. The experiment demonstrated that students were highly engaged, curious, and motivated to test new sales scenarios, particularly by gaining initial experience in handling challenging interactions with unfriendly customers.
While the study did not reveal statistically significant differences between conditions, it identified important trends and insights that can guide the design of future simulations. By prioritizing positive user interactions, improving realism, and addressing technical challenges, virtual environments can be enhanced to better meet user needs. Future research should delve deeper into the complex interactions between contextual factors, user perceptions, and system design to further refine the development of effective and engaging virtual simulations.



\subsection{Limitations and Future Work}
This study has several limitations that should be addressed in future research. First, the lack of statistically significant differences across conditions may be due to a limited sample size, which could reduce statistical power. Future studies should consider larger sample sizes to better detect subtle effects. Second, while the quantitative measures provided valuable insights, they may not fully capture participants’ subjective experiences. Integrating more qualitative methods, such as in-depth interviews or focus groups, could provide a richer understanding of user perceptions. Third, the conversation with the avatars was relatively narrow, following a premade script. In the near future, the integration of LLM will be explored to foster the training effectiveness. Our hypothesis is that it will allow participants to engage in dynamic, context-dependent dialogues within realistic settings, enabling them to respond flexibly to customer inquiries and objections. Additionally, the integration of multimodal feedback mechanisms, which extend beyond visual and auditory feedback to include emotional response tracking, could further enhance the training's effectiveness. 


\begin{credits}
\subsubsection{\ackname} This work was supported by the European Union's Horizon Europe programme under grant number 101092875 ``DIDYMOS-XR'' (https://www.didymos-xr.eu).

In this paper, we used Overleaf’s built-in spell checker, the current version of ChatGPT (GPT 4.0), and Grammarly. These tools helped us fix spelling mistakes and get suggestions to improve our writing. If not noted otherwise in a specific section, these tools were not used in other forms.
\end{credits}

%
%
%
%

\end{document}